\documentclass[onecolumn,preprint,superscriptaddress]{revtex4-1}
\usepackage{bm}
\usepackage[colorlinks=true,linkcolor=blue,citecolor=blue]{hyperref}
\usepackage{times}
\usepackage{amsmath}
\usepackage{amssymb}
\usepackage{amsthm}
\usepackage{amsfonts}
\usepackage{enumerate}
\usepackage{latexsym}
\usepackage{ifpdf}
\usepackage{natbib}
\usepackage{psfrag}
\newcommand{\beq}{\begin{equation}}
	\newcommand{\eeq}{\end{equation}}
\usepackage{graphicx}
\usepackage{makeidx}
\usepackage{color}
\usepackage{array}
\usepackage{multirow}
\usepackage{siunitx}
\usepackage{}

\begin{document}
	
	\title{High-speed sensing of RF signals with phase change materials}
	
	\author {Ranjan Kumar Patel}
	\altaffiliation{Contributed equally}
	\email{ranjan.patel@rutgers.edu}
	\affiliation  {Department of Electrical and Computer Engineering, Rutgers University, Piscataway, NJ 08854, USA}
	\author {Yifan Yuan}
	\altaffiliation{Contributed equally}
	\affiliation  {Department of Electrical and Computer Engineering, Rutgers University, Piscataway, NJ 08854, USA}
	\author {Ravindra Singh Bisht}
	\affiliation  {Department of Electrical and Computer Engineering, Rutgers University, Piscataway, NJ 08854, USA}
	\author {Ivan Seskar}
	\affiliation  { WINLAB, Rutgers University, North Brunswick, NJ 08902, USA}
	\author {Narayan Mandayam}
	\affiliation  { WINLAB, Rutgers University, North Brunswick, NJ 08902, USA}
	\author {Shriram Ramanathan}
	\email{shriram.ramanathan@rutgers.edu}
	\affiliation  {Department of Electrical and Computer Engineering, Rutgers University, Piscataway, NJ 08854, USA}

	\begin{abstract}
		RF radiation spectrum is central to wireless and radar systems among numerous high-frequency device technologies. Here, we demonstrate sensing of RF signals in the technologically relevant 2.4 GHz range utilizing vanadium dioxide (VO$_2$), a quantum material that has garnered significant interest for its insulator-to-metal transition. We find the electrical resistance of both stoichiometric as well as off-stoichiometric vanadium oxide films can be modulated with RF wave exposures from a distance. The response of the materials to the RF waves can be enhanced by either increasing the power received by the sample or reducing channel separation. We report a significant $\sim$ 73 $\%$ drop in resistance with a 5 $\mu$m channel gap of the VO$_2$ film at a characteristic response time of 16 microseconds. The peak sensitivity is proximal to the phase transition temperature boundary that can be engineered via doping and crystal chemistry. Dynamic sensing measurements highlight the films' rapid response and broad-spectrum sensitivity. Engineering electronic phase boundaries in correlated electron systems could offer new capabilities in emerging communication technologies.
		
	\end{abstract}
	
	\maketitle
	\section{Introduction}
	
	Sensing of RF signals and wireless spectrum has increasingly become essential for a wide variety of uses. These range from the classical need for spectrum sensing in cognitive radio (CR) ecosystems to opportunistically use the available spectrum to the more generic uses of RF signals for a variety of applications supported by the Internet of Things (IoT) ecosystems. In CR scenarios, spectrum sensing has been used to create radio maps that allow secondary spectrum users to exploit available spectrum holes and peacefully coexist with primary (incumbent) users of spectrum~\cite{Akyildiz2006p2127}. In IoT ecosystems, RF signal sensing has been used for a variety of applications, e.g., environmental monitoring, healthcare, advanced manufacturing~\cite{Lubna:2022}, to name a few. Further, the developments in the evolution of 6G wireless technologies have underscored the importance of sensing by seeking to integrate communications and sensing in a joint framework~\cite{Tan:2021p1}.

	The need for RF sensing without necessarily using complex (and often frequency-specific) signal processing is attractive. Further, having such sensing accomplished at high speeds is particularly relevant for not only supporting low latency communications but also for enabling follow-up actions related to possibly network security and control. While the sensing methodology envisioned here is applicable even in the far-field, near-field sensing is becoming increasingly relevant with the emergence of the IoT ecosystem, with high densities of devices in close geographical proximity to each other for various machine-to-machine communications scenarios, including sensors in body and personal area networks. Motivated by the above, we investigate RF sensing utilizing a novel materials-based sensing approach in the near-field that can be used seamlessly in a wide variety of scenarios.

	Materials exhibiting electronic phase transitions are well-suited for sensing applications. Vanadium dioxide (VO$_2$), a prototypical quantum material characterized by its insulator-to-metal transition (IMT) near room temperature, has been explored as switches~\cite{Chae:2005p76,Anagnostou:2020p58,Ghazikhanian:2023p123505} and sensors for chemical, thermal, and terahertz detection~\cite{Kim:2007p023515,Strelcov:2009p2322,Qaderi:2023p34,Qaderi:2022p1}. The IMT characteristics can be further modified by utilizing an off-stoichiometric vanadium oxide compound, denoted as VO$_x$~\cite{Zhang:2017p034008,Zhang:2017p27135,Ramirez:2015p205123}. Simultaneously, the V:O materials family is recognized for its efficacy as a bolometer~\cite{Wang:2012p3,Wang:2023p168295,Wang:2021p70}, exemplifying the versatile applications of such materials in sensing technologies. Here, we demonstrated the effect of RF waves on VO$_2$ and off-stoichiometric VO$_x$ films on $c$-Al$_2$O$_3$ (0001) substrate by varying different parameters, such as the temperature, frequency, device geometry, distance of the film from the RF antenna, gain and power of the RF waves. We particularly focus on the 2.4 GHz frequency range due to its importance in wireless communications. Our findings suggest that the studied materials can be promising candidates for RF sensors, potentially contributing to advancements in future Wi-Fi technologies.

	\section{Experimental details}
	
	VO$_2$ and off-stoichiometric VO$_x$ films of thickness $\sim$40 nm were grown on $c$-axis (0001) sapphire (Al$_2$O$_3$) substrates by a radio-frequency (RF) magnetron sputtering (AJA International) system~\cite{Zhang:2017p034008}. A ceramic V$_2$O$_5$ target of 99.9 $\%$ purity was used with 100 W RF power. The V$_2$O$_5$ target was pre-sputtered for 5 minutes before the deposition. During deposition, the pressure was maintained at 5 mTorr by introducing 49.5 SCCM (standard cubic centimeters per minute) Ar and 0.5 SCCM O$_2$-Ar (10$\%$-90$\%$) mix gases for VO$_2$ growth, whereas 49.9 SCCM Ar and 0.1 SCCM O$_2$ gases were used for VO$_x$ growth. The substrate temperature was 650$^{\circ}$C, and the substrate holder was rotated at 40 rpm during the growth to maintain the homogeneity of the sample. Post deposition, the substrate was cooled down to room temperature at the growth pressure. Pt and Ni electrodes were deposited (using sputtering) at room temperature on VO$_2$ and VO$_x$  films using a shadow mask, respectively, for electrical measurements across millimeter scale junctions. To investigate micro-scale junctions, VO$_2$ devices with 5 $\mu$m, 15 $\mu$m, 25 $\mu$m, and 30 $\mu$m separations between the electrodes were fabricated through photolithography using a photoresist of AZ 1518 as the masking layer for the process. Heidelberg MLA150 Maskless Aligner was used to write the electrode pattern~\cite{Deng:2023p4838}. A 100-nanometer thick layer of Pt was deposited through e-beam evaporation and subsequently lifted off by using PG-Remover at 80$^{\circ}$C.
	
	The X-ray diffraction (XRD) patterns of the substrate and as-deposited films were recorded by using a lab-based Panalytical Xpert diffractometer with a Cu K$_{\alpha}$ radiation (1.5406 \AA) source. Rutherford backscattering spectroscopy (RBS) measurements were performed to estimate the stoichiometry by using a 1.7 MV IONEX tandem accelerator with a 2.3 MeV He$^{2+}$ ion beam of diameter 2 mm. The scattering angle of the detector was 163 degrees, and the resolution of the detector was 18 keV. The RBS data were analyzed by using the SIMNRA program~\cite{Mayer:2002p177}. The $dc$ transport measurements were performed on a probe station by using a Keithley 2635A source meter, and the temperature was controlled by using a Quiet CHUCK DC Hot Chuck system (see Fig. S1 of Supplementary Materials for experimental setup). The electrical contacts were made by contacting the micromanipulator tips of the probe station to the metal electrodes directly. The resistance was measured by taking the slope of the $IV$ data taken between -0.05 V and +0.05 V in a two-probe configuration. The time-resolved measurements were performed by using a Keithley 2461 source meter.
	
	In the RF measurement setup, an X86-based Software Defined Radio (SDR) platform (Quad-core i7 embedded PC) was used and equipped with a B210 USRP (Universal Software Radio Peripheral) and a directional antenna. The USRP, interfaced with the SDR platform, utilized User Hardware Driver (UHD) tools for the precise generation and control of RF waveforms. To augment the system, it was connected to the output of the USRP, significantly enhancing the emitted RF signal strength. This setup was instrumental in effectively exciting VO$_2$ and VO$_x$ films on $c$-Al$_2$O$_3$ substrates. The waveform used was a narrow band sine wave, with the gain values listed in the paper corresponding to the UHD tx$\_$waveform utility command line gain argument. The directional antenna, in conjunction with the amplified signal from the power amplifier, focused the RF energy onto the samples in the lab, as shown in Fig. S1 of Supplementary Materials. The software-based waveform configuration enabled precise control over key parameters, such as frequency and power. We calibrated our setup with a spectrum analyzer in the COSMOS testbed environment at WINLAB~\cite{Raychaudhuri:2020p13}, measuring the received power by the films at various distances from the SDR setup and with different gain values (see Fig. S2 of Supplementary Materials for the calibration setup). All the measurements were done in the near-field regime.

	\section{Results and Discussion}

\begin{figure*}
		\vspace{-0pt}
		\includegraphics[width=\textwidth] {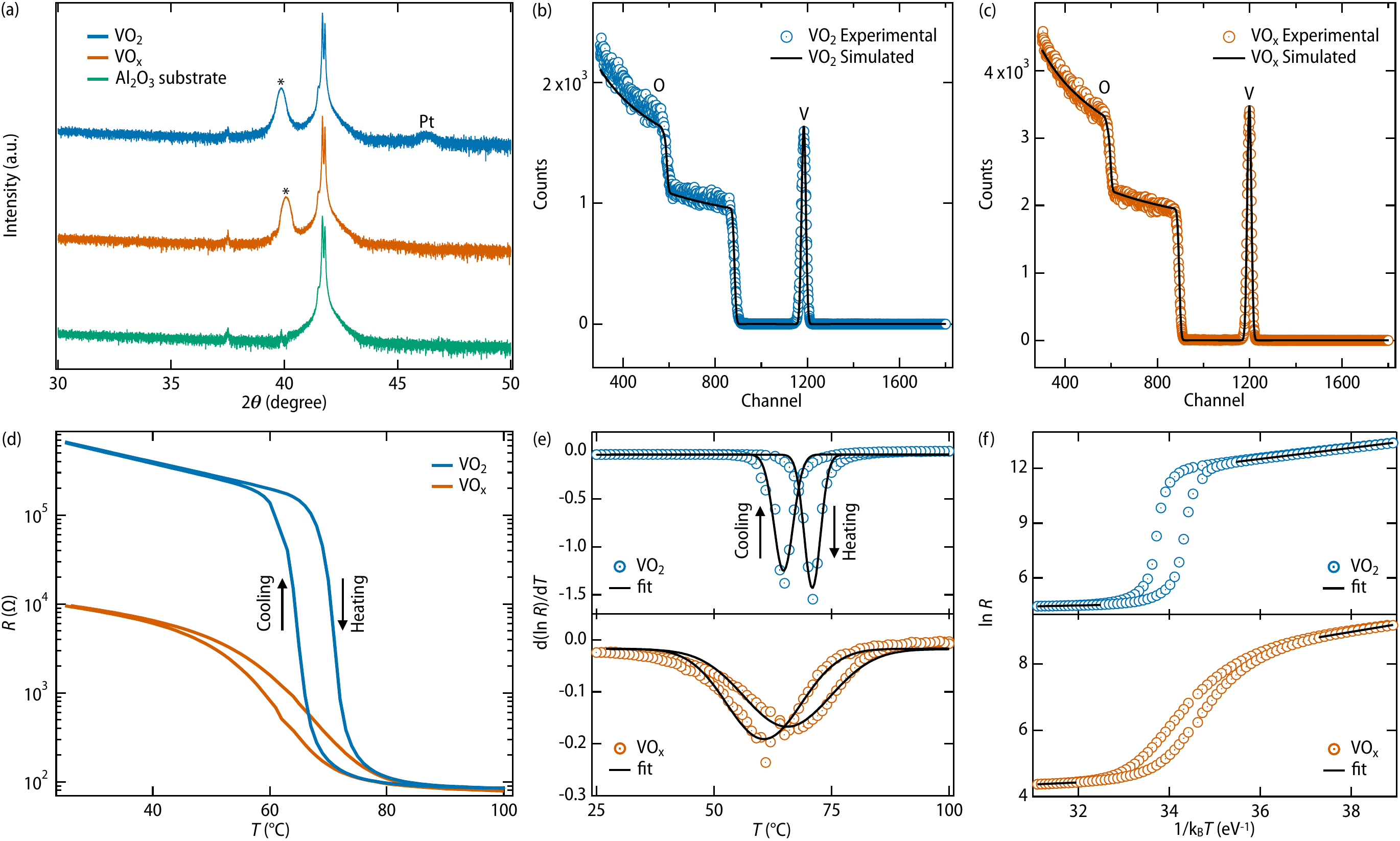}
		\caption{\label{Fig1} {(a) 2$\theta$-$\omega$ XRD scan for VO$_2$ and VO$_x$ films. (b)-(c) Experimental and simulated Rutherford backscattering spectroscopy data for VO$_2$ and VO$_x$ films, respectively. (d) Resistance as a function of temperature of the VO$_2$ and VO$_x$ films. (e) Gaussian fitting of the differential curves of resistance vs. temperature to extract the IMT for the VO$_2$ and VO$_x$ films. (f) Estimation of activation energy of the VO$_2$ and VO$_x$ film by linear fitting of the ln(R) vs. 1/$k_{B}T$ plots in the  cooling run.}}
		
	\end{figure*}

	In order to check the structural quality of the VO$_2$ and VO$_x$ films grown on sapphire substrate by RF sputtering, we have recorded 2$\theta$-$\omega$ diffraction scan [Fig.~\ref{Fig1}(a)]. For comparison, the XRD pattern of the single crystal sapphire substrate was also measured and shown in Fig.~\ref{Fig1}(a). The XRD scan of the VO$_2$ and VO$_x$ films consists of a broader film peak (marked by $\ast$) and a sharp substrate peak (2$\theta$ = 41.68$^{\circ}$)~\cite{Cui:2011p041502}, where the substrate peak arises due to the (0006) reflection of the $c$-Al$_2$O$_3$. For the VO$_2$ film, a well-defined XRD peak corresponding to the monoclinic (020) phase~\cite{Cui:2011p041502} was observed at 2$\theta$ = 39.86$^{\circ}$. From this data, the lattice distance was estimated ($b_m$ = 4.52 \AA), which is similar to the respective literature value of expected monoclinic VO$_2$~\cite{Zhou:2012p074114}. For the off-stoichiometric sample (VO$_x$), the XRD peak was observed at 2$\theta$ = 40.08$^{\circ}$, which indicates a value of $b_m$ = 4.5 \AA. This slight shift of the XRD peak is related to the oxygen non-stoichiometry in the VO$_x$ film~\cite{Rampelberg:2015p11357,Ding:2023p3778}. Apart from this peak shift, no new diffraction peaks are observed. Additionally, to estimate the elemental composition, we performed Rutherford backscattering spectroscopy (RBS) experiments. In RBS, the qualitative determination of the areal density of the elements is possible by analyzing the intensity and energy of the backscattered He$^{2+}$ ions from the sample within 1-2$\%$ accuracy~\cite{Feldman1986}. Figures~\ref{Fig1}(b)-(c) show the RBS analysis data of VO$_2$ and VO$_x$ films, respectively, and the values obtained from the analysis are summarized in Table~\ref{tab:Table1}. For VO$_2$ and VO$_x$ films, the ratio of the spatially averaged areal density of vanadium to oxygen (V:O) are estimated as 1:2 and 1:1.7, respectively.

	\begin{table*}
		\caption{\label{tab:Table1}Summary of RBS and transport data}
		\begin{ruledtabular}
			\begin{tabular}{ccccccc}
				\mbox{sample}&\mbox{V}&\mbox{O}&\mbox{$T^h_\mathrm{IMT}$}&\mbox{$T^c_\mathrm{IMT}$}&\mbox{E$^c_a$ (I)}&\mbox{E$^c_a$ (M)}\\
				
				&\mbox{($\times$10$^{15}$ atoms/cm$^2$)}&\mbox{($\times$10$^{15}$ atoms/cm$^2$)}&\mbox{($^{\circ}$C)}&\mbox{($^{\circ}$C)}&\mbox{(meV)}&\mbox{(meV)}\\
				\hline \\
				VO$_2$ & 101.9 & 204.1 & 70.6  & 64.1   & 302.1$\pm$1.7 & 52.5$\pm$3 \\
				VO$_x$ & 103.6 & 176.4 & 64.3  & 59.4   & 227.7$\pm$4.9 &  62.3$\pm$2.2\\

			\end{tabular}
		\end{ruledtabular}
	\end{table*}

	Following the structural measurements, we have investigated the electrical transport properties of the films. As reported earlier~\cite{Zhang:2017p034008,Cui:2011p041502,Zhou:2012p074114,Yang:2011p033725,Wong:2013p74}, VO$_2$ film on Al$_2$O$_3$ substrate undergoes an insulator-to-metal transition (IMT) accompanied by a structural change from monoclinic to tetragonal rutile structure upon increasing the temperature. Figure~\ref{Fig1}(d) shows the temperature-dependent resistance of VO$_2$ and VO$_x$ films on $c$-Al$_2$O$_3$ substrate. For the stoichiometric VO$_2$ sample, the resistance changes abruptly around four orders of magnitude, which verifies the high quality of the film. On the other hand, VO$_x$ film undergoes a resistance change of around two orders of magnitude. This is characteristic of oxygen-deficient vanadium oxides that show suppressed insulating state resistance and transition temperature. The IMT temperature was calculated by plotting the d(ln $R$)/d$T$ as a function of temperature and taking the maximum magnitude after fitting with a Gaussian curve [Fig.~\ref{Fig1}(e)]. The IMT of the VO$_2$ in heating ($T^h_\mathrm{IMT}\sim$ 70.6$^{\circ}$C) and cooling runs ($T^h_\mathrm{IMT}\sim$ 64.1$^{\circ}$C) are higher compared to the bulk VO$_2$ ($T^h_\mathrm{IMT}\sim$ 68$^{\circ}$C) and is related to the tensile strain~\cite{Zhou:2012p074114,Muraoka:2002p583,Chen:2010p211905}. Again, for the off-stoichiometric VO$_x$ sample, $T^h_\mathrm{IMT}$ and $T^c_\mathrm{IMT}$ are calculated as 64.3$^{\circ}$C and 59.4$^{\circ}$C, respectively, which are lesser compared to the stoichiometric VO$_2$ sample. Furthermore, the activation energy (E$_a$) analysis for both VO$_2$ and VO$_x$ films in the metallic and insulating phases is shown in Fig.~\ref{Fig1}(f), which is calculated by plotting ln $R$ as a function of 1/k$_BT$ and taking the slope~\cite{Yang:2011p033725}. The activation energy for VO$_2$ sample in the metallic and insulating phases are found to be 52.5$\pm$3 meV and 302.1$\pm$1.7 meV, respectively, in the cooling run. A similar analysis on the VO$_x$ sample yields E$^c_a$=62.3$\pm$2.2 meV and 227.7$\pm$4.9 meV for the respective metallic and insulating phases.
	
	\begin{figure*}
		\vspace{-0pt}
		\includegraphics[width=\textwidth] {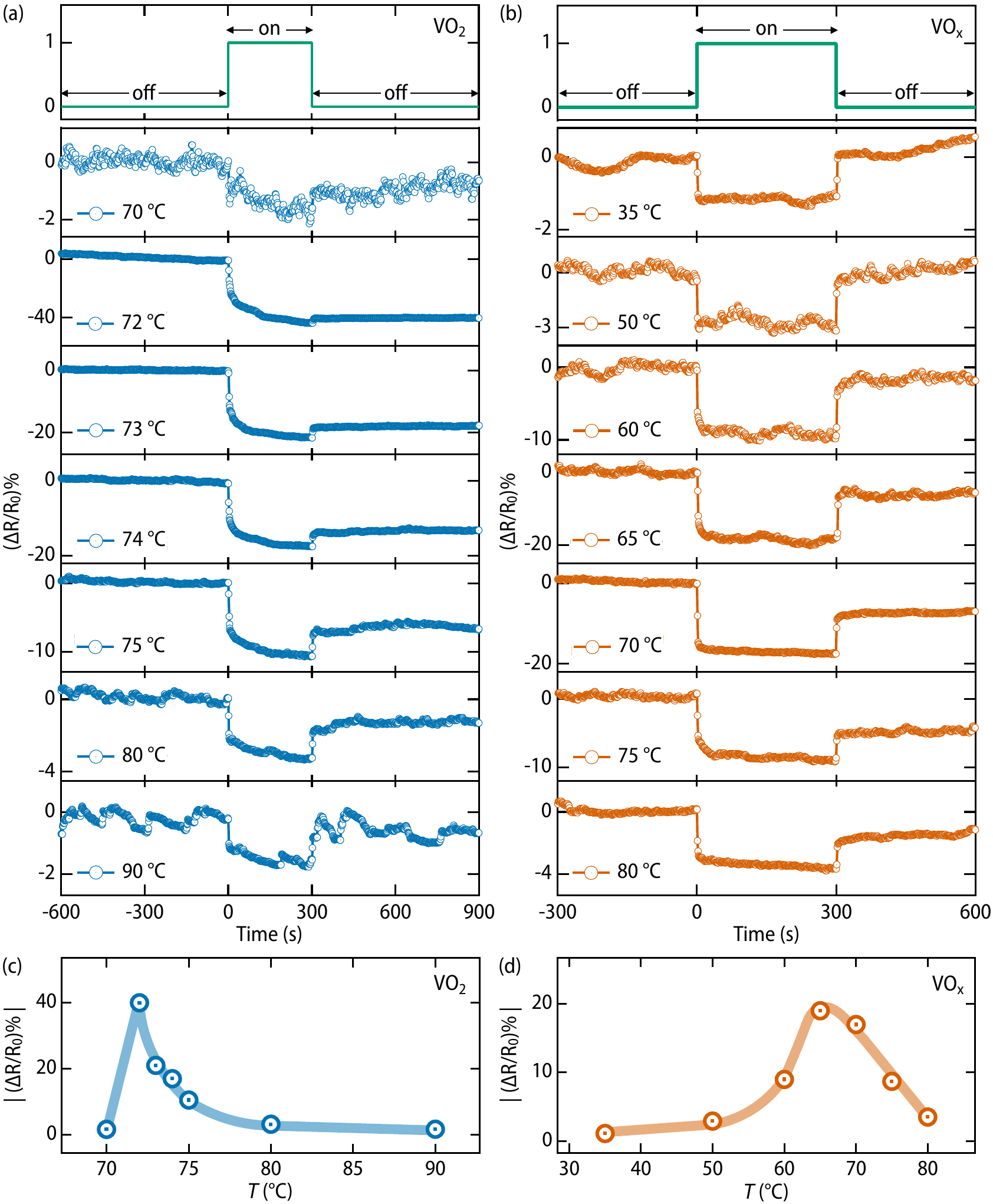}
		\caption{\label{Fig2} {(a)-(b) Relative percentage change in resistance under irradiation of RF signal (2.4 GHz with 100 gain for 5 min at 8 cm distance) at different temperatures of the VO$_2$ and VO$_x$ films, respectively. The top panels of (a) and (b) show the schematics representing the on/off-state of the RF signal with time. (c)-(d) Temporal evolution of the maximum relative percentage change in resistance of the VO$_2$ and VO$_x$ films, respectively. }}
	\end{figure*}
	
	After probing the structural and electronic properties of the stoichiometric and off-stoichiometric samples, we delved into the impact of radio frequency (RF) wave exposure on these films. The upper panels of Figs.~\ref{Fig2}(a)-(b) show the on/off-state of the RF signal. The samples were heated to the desired temperature for each measurement. Once the temperature stabilized, the system was driven out of equilibrium by concentrating a 2.4 GHz RF wave with 100 gain from a distance of 8 cm onto the samples for 5 minutes during the on-state. Following each set of sensing measurements at a specific temperature, the films were ramped to 100$^{\circ}$C to remove any persistent conductivity or history effects. Subsequently, the samples were cooled down to room temperature and then heated to the desired temperature for the next measurement. Figures~\ref{Fig2}(a)-(b) show the relative percentage change in resistance (RPCR), calculated using the formula ((R-R$_0$)/R$_0$)*100 (R$_0$ is the initial resistance when the RF signal was turned off) as a function of time at different temperatures of the VO$_2$ and VO$_x$ samples, respectively. Firstly, a drop in resistance was observed as soon as the RF signal was activated, making the material more conducting. It is worth noting that the RF wave with a frequency of 2.4 GHz corresponds to a energy of 9.93 $\mu$eV, which is not sufficient for the excitation of electrons from the valence band to the conduction band as the band gap of VO$_2$ is around 0.7 eV~\cite{Yang:2011p033725}. Therefore, this decrease in resistance induced by the RF wave may be due to the selective excitation of trapped electrons to the conduction band. Prior studies have demonstrated an IMT phase transition in VO$_2$ coplanar waveguides induced by intense high frequency radiation~\cite{Liu2012:345,Ha:2014p575}. The decrease in resistance caused by RF waves may also occur due to alternative mechanisms, such as influencing the formation of conductive filaments within the VO$_2$ film~\cite{Qaderi:2023p34,Qaderi:2022p1,Duchene:2003p115,Lee:2008p443} or through the liberation of Poole-Frenkel electrons, where the electric field reduces the potential barrier, leading to a slight elevation in the carrier density~\cite{Liu2012:345,Simmons:1967p657}. Moreover, the reduction in the RCPR due to the influence of RF waves can be anticipated through a straightforward application of the Joule heating mechanism~\cite{Liu2012:345,Sun:2021p035302}. In this latter scenario, RF waves locally heat the film, causing a small temperature rise and subsequently making the sample more conductive. Secondly, there is a sharp increase in resistance immediately when the RF signal is turned off, which suggests a different mechanism potentially distinct from only Joule heating as the resistance typically recovers gradually in the case of heat dissipation~\cite{Fleck:2016p064015,Lu:2015p37}. To check whether this resistance change with the RF signal is related to the material's structure, we conducted X-ray diffraction (XRD) measurements after shining 2.4 GHz RF waves on a VO$_2$ sample for 5 minutes. However, there was no observable change in the film peak (see Fig. S3 of Supplementary Materials). Thirdly, when the samples were in either the insulating or metallic state, the resistance recovered back to the original resistance (R$_0$), whereas the resistance did not fully recover to the original value after removing the RF signal when the samples were in the hysteresis region. This behavior can be well explained by the hysteresis effect, where the resistance does not return to its initial value if the temperature is ramped up (down) and down (up) only within the hysteresis region~\cite{Gomez:2018p8479,Gurvitch:2010p395}. The abrupt change of channel resistance with RF waves at different temperatures is fascinating as this can be exploited as a design parameter in sensing. Figures~\ref{Fig2}(c)-(d) illustrate the RPCR as a function of measured temperatures for the VO$_2$ and VO$_x$ samples, respectively. Interestingly, the maxima of the curves intriguingly close to the $T^h_\mathrm{IMT}$, which suggests maximum changes in resistance approaching the percolation threshold~\cite{Koughia:2020p063401,Annasiwatta:2016p6905, Wang:2015p085150}. In future, in-operando microscopy techniques such as scanning microwave impedance microscopy (SMIM) and scanning near-field infrared microscopy (SNIM) coupled with RF exposure could enable a comprehensive understanding of the microscopic mechanisms involved and could form the subject of further studies.

	\begin{figure*}
		\vspace{-0pt}
		\includegraphics[width=0.85\textwidth] {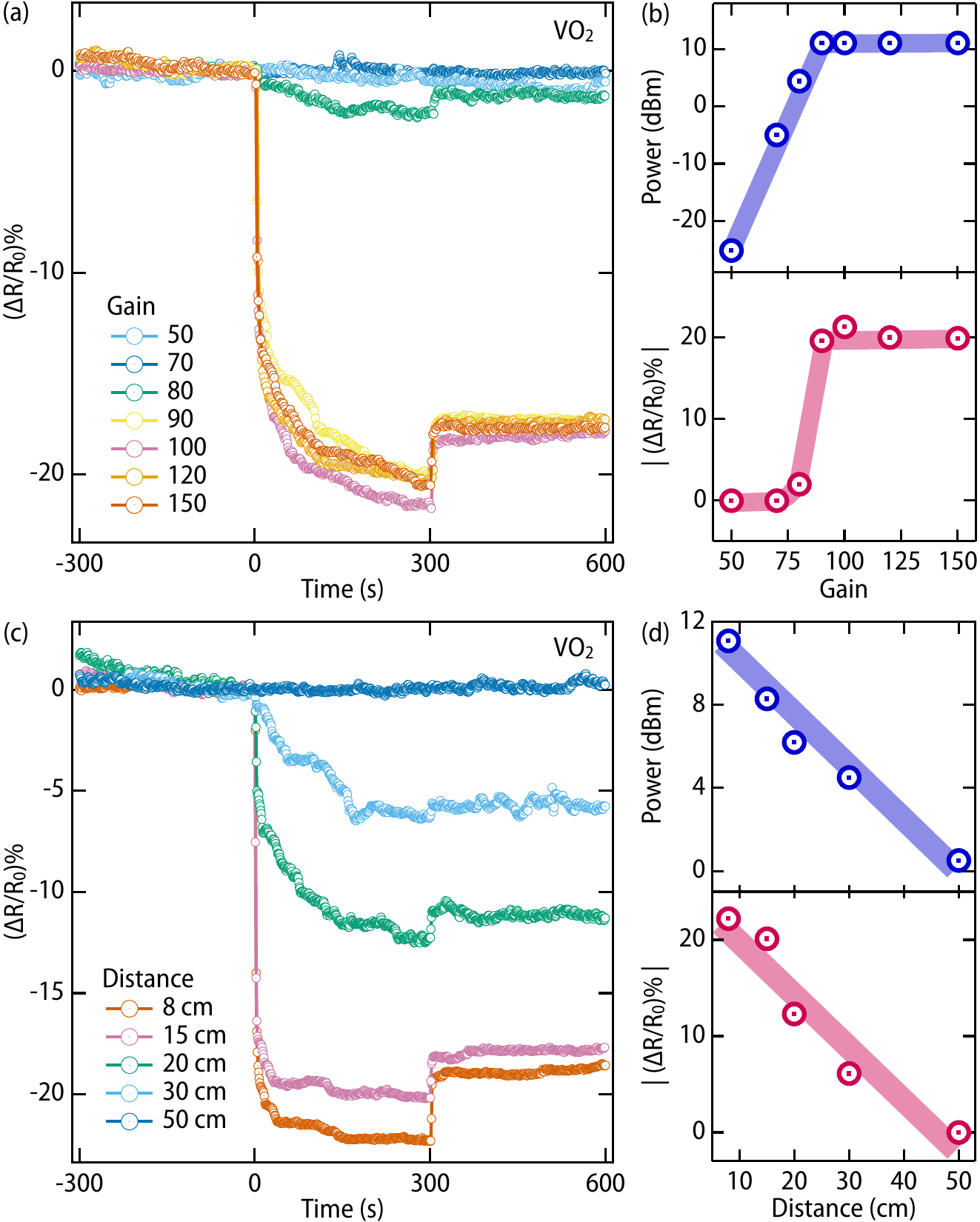}
		\caption{\label{Fig3} {(a) Relative percentage change in resistance (RPCR) under RF signal (2.4 GHz at 73$^{\circ}$C for 5 min at 8 cm distance) at different gain values of the VO$_2$ film. (b) The power received and the maximum relative percentage change in resistance change with the gain. (c) RPCR of the VO$_2$ film under RF signal (2.4 GHz with a gain of 100 at 73$^{\circ}$C for 5 min) from different distances. (d) The power received and the maximum relative percentage change in resistance change with distances from the RF antenna. }}
	\end{figure*}
	
	To check the reproducibility, a 2.4 GHz RF wave with 100 gain was concentrated on a VO$_x$ sample for 5 minutes from a distance of 8 cm at 73$^{\circ}$C. This procedure was then repeated for the same sample and also for a different sample (see Fig. S4 of Supplementary Materials). The RPCR for all these cases are found to be almost identical. Additionally, we explored the influence of RF waves on the sample using frequencies other than the technologically important 2.4 GHz used in wireless communications such as Wi-Fi and cellular applications. We observed similar conductance modulation effects on the VO$_2$ film (see Fig. S5 of Supplementary Materials), suggesting that the conductance modulation phenomenon is not exclusive to specific frequencies but holds true across a range of RF waves. This versatility in responsiveness to diverse frequencies positions VO$_2$ as a potential candidate for applications in the evolving landscape of communication technologies.

	After establishing the effect of RF waves on VO$_2$ and VO$_x$ samples at different temperatures, we explored the effect of RF signal strength by varying the power. Figure~\ref{Fig3}(a) shows the RPCR of the VO$_2$ sample as a function of time when radiated with a 2.4 GHz RF signal (for 5 minutes at a distance of 8 cm) at 73$^{\circ}$C with different gain values. The power received at the sample surface corresponding to the gain and the maximum relative percentage change in resistance with gain is shown in Fig.~\ref{Fig3}(b). Notably, the maximum RPCR remains minimal at lower gains (50 and 60), increases progressively with higher gain values, and stabilizes when the gain exceeds 80. It's worth noting that the RPCR is directly proportional to the power received by the sample at a constant temperature. Furthermore, Fig.~\ref{Fig3}(c) shows the RPCR as a function of time, while the RF wave was concentrated on the VO$_2$ film from different distances between the RF antenna and the sample. As can be seen from the graph, the RPCR decreases with increasing distance between the RF source and the sample. This can be understood by considering the decrease of power with increasing distance. To validate this, we measured the power at the corresponding distances and found that the power indeed decreases as the distance increases [Fig.~\ref{Fig3}(d)]. Therefore, the observed changes in RPCR concerning gain and distance can be attributed to the power received by the sample.

\begin{figure*}
		\vspace{-0pt}
		\includegraphics[width=0.7\textwidth] {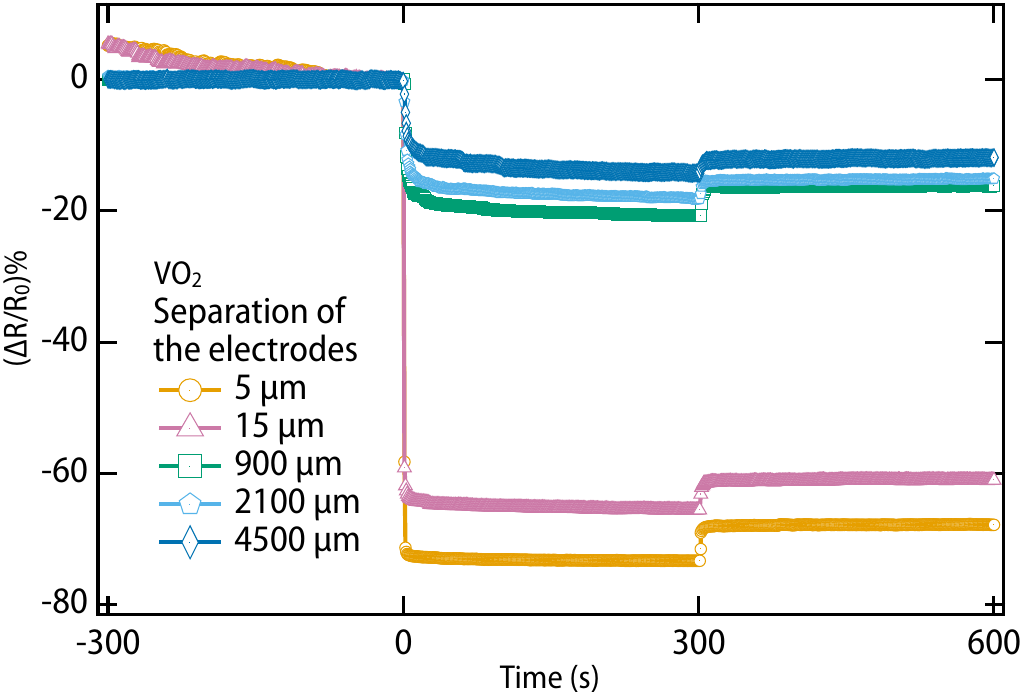}
		\caption{\label{Fig4} {Relative percentage change in resistance with a 2.4 GHz RF signal (with a gain of 100 at 73$^{\circ}$C for 5 min at a distance of 8 cm) of VO$_2$ films with different separations of electrodes.}}
	\end{figure*}

	Next, we examined the effect of RF waves on sensor channel dimensions by varying the spacing between the metal electrodes on the VO$_2$ sample. Figure~\ref{Fig4} shows that an increase in the distance between the electrodes (or sample size) results in a decrease in the RPCR when a 2.4 GHz RF wave with a gain value of 100 from a distance of 8 cm was turned on at a sample temperature of 73$^{\circ}$C. The metal electrodes were deposited using a shadow mask for the VO$_2$ film with a channel separation of 900 $\mu$m, 2100 $\mu$m, and 4500 $\mu$m. Further, we used another VO$_2$ film where the electrodes were patterned by lithography with much smaller separations to enhance the RPCR. The relative percentage change in the resistance were found to be $\sim$ 73 $\%$ and 65 $\%$, when the electrode separations were 5 $\mu$m and 15 $\mu$m, respectively. The enhancement in sensitivity by reducing the channel separation points towards further opportunities for the optimization of sensors for practical technologies.
	
	\begin{figure*}
		\vspace{-0pt}
		\includegraphics[width=0.9\textwidth] {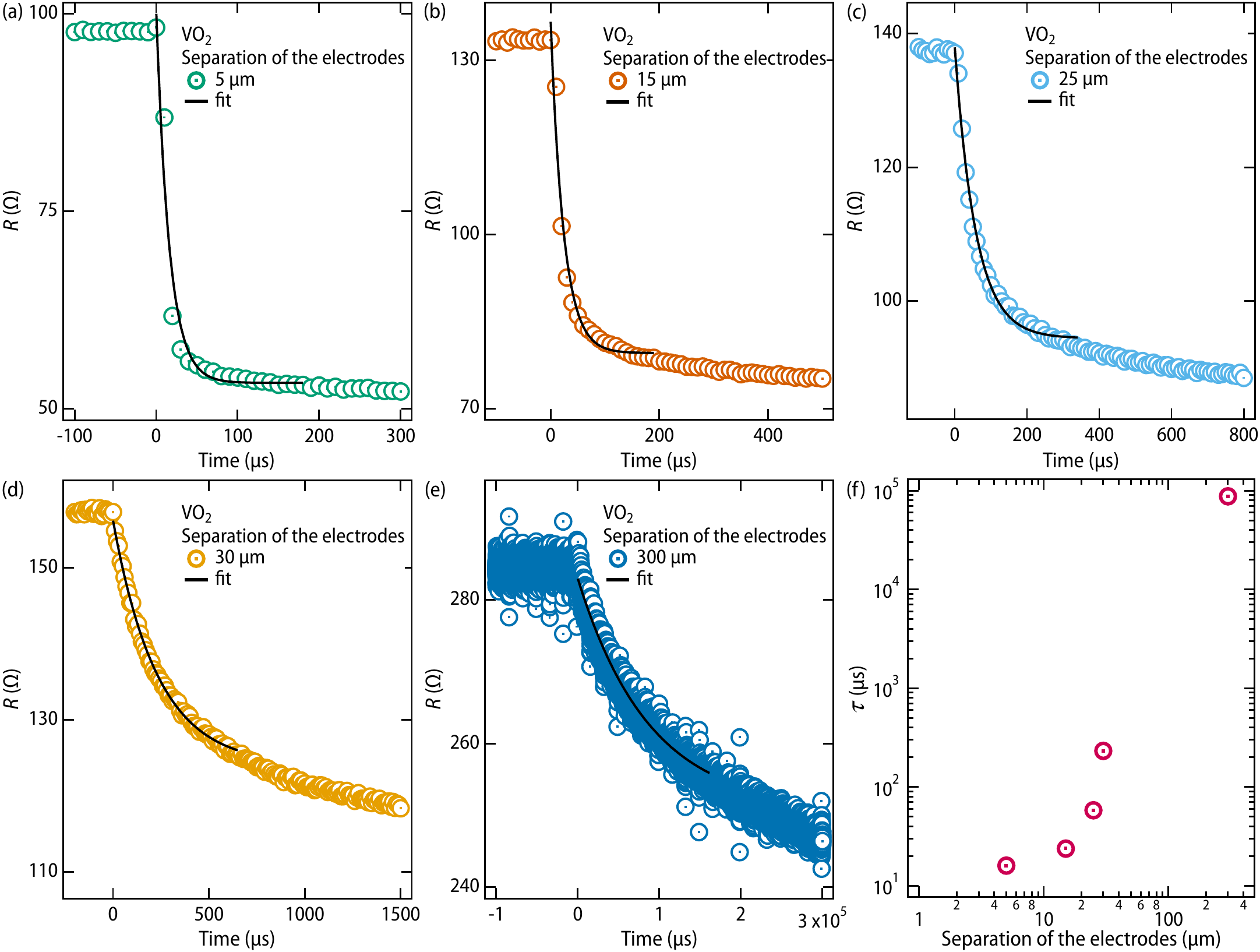}
		\caption{\label{Fig5} {(a)-(e) Resistance as a function of time in microsecond time scale with a 2.4 GHz RF signal (with a gain of 100 at 73$^{\circ}$C at a distance of 8 cm) of VO$_2$ films with different separations of electrodes. The black curves are the exponential fit to the data to extract the time constant. (f) Time constant as a function of the separation of the electrodes of the VO$_2$ film.}}
	\end{figure*}
	
	To elucidate the dynamics associated with the reduction in resistance induced by RF waves, we conducted time-resolved measurements, capturing resistance values at intervals of 10 microseconds. Figures ~\ref{Fig5}(a)-(e) show the temporal evolution of resistance drop upon the activation of RF waves, focusing on various channel separations within the VO$_2$ film (see Fig. S6 of Supplementary materials for the evolution of resistance for a longer time scale). It is noteworthy to highlight that the response time scales varies with channel dimensions. The drop in resistance exhibits a distinct profile, characterized by a sharp and rapid reduction in the case of shorter channels, transitioning to a more gradual descent as the channel separation is increased. To understand the time constant related to the drop of the resistance with RF wave, we fitted the resistance ($R$) vs. time ($t$) curve (immediately following the RF exposure) with the equation, $R=a+be^{-t/\tau}$, where $a, b$ are constants and $\tau$ is the time constant [black curves in Figs.~\ref{Fig5}(a)-(e)]. The time constant increases with the increase of the channel separation of the VO$_2$ film [Fig.~\ref{Fig5}(f)]. This insight points towards the intricate interplay between channel dimensions and the manifestation of RF-induced effects, particularly in the context of conducting filament formation within the VO$_2$ film that has previously been ascribed as a possible mechanism for reduction in resistance of the vanadium oxide films exposed to THz fields~\cite{Qaderi:2023p34,Qaderi:2022p1}. The high speed response could potentially be enhanced further by reducing channel separation through the application of advanced techniques such as e-beam lithography or other similar methods.

	\section{Conclusions}

	To summarize, we have grown VO$_2$ and off-stoichiometric VO$_x$ films on $c$-Al$_2$O$_3$ substrates and studied their structural and electrical properties by XRD, RBS, and transport measurements. Further, we have investigated the effect of 2.4 GHz radiation on these samples by varying the temperature, frequency, gain, distance, power, and sample size. Interestingly, the application of the RF wave makes the samples more conducting and opens up avenues for applications as RF sensors. The relative percentage change in resistance scales with decreasing channel separation, reaching a value of $\sim$ 73 $\%$ for the 5 $\mu$m channel gap. We found that the samples are most sensitive proximal to the transition boundary, and this offers a path to devices that can respond at various temperatures by varying crystal stoichiometry and doping. The time-resolved RF measurements suggest the rapid response of the film in microsecond time scales upon the incidence of RF waves. Furthermore, the influence of RF waves is detectable across a broad spectrum of microwave frequencies, offering potential applications in future communications technologies.

	\section{Acknowledgement}
	
	The authors acknowledge Analytical Characterization Laboratories and Laboratory for Surface Modification Facilities at Rutgers University for XRD and RBS measurements, respectively. We acknowledge Haoming Yu for the fabrication through photolithography of the VO$_2$ sample. We are grateful to Prof. Michael Gershenson and Plamen Kamenov for the use of the sputtering system. We acknowledge Air Force Office of Scientific Research AFOSR FA9550-23-1-0215 that supported the film growth.

%

%

\end{document}